\documentstyle[twoside,fleqn,espcrc2]{article}
\input epsf

\title{Chiral perturbation theory for $K^+ \rightarrow \pi^+ \pi^0$ 
       decay in the continuum and on the lattice}
       
\author{Maarten Golterman$^{\rm a}$
        and 
        Ka Chun Leung\address{Department of Physics, Washington University,
        St. Louis, MO 63130, USA.}
        \thanks{presenter at conference}
        }

\begin{document}

\begin{abstract}
We use one-loop chiral perturbation theory (ChPT) to compare lattice 
results for the $K^+ \rightarrow \pi^+ \pi^0$ decay amplitude with 
the experimental value. Three systematic effects: quenching, 
finite-volume effects, and the use of unphysical values of quark masses 
and pion external momenta can be investigated. We find that the corrections 
help in explaining the discrepancy between lattice and experimental results. 
We also discuss the relation to $B_K$.
\end{abstract}

% typeset front matter (including abstract)
\maketitle

\section{Basic theory}
In the Standard Model,
the decay $K^+\to\pi^+\pi^0$ is induced by the four-fermion operator $O_4 =
{\bar s}_L \gamma^\mu d_L \ {\bar u}_L \gamma_\mu u_L
+{\bar s}_L \gamma^\mu u_L \ {\bar u}_L \gamma_\mu d_L
-{\bar s}_L \gamma^\mu d_L \ {\bar d}_L \gamma_\mu d_L$. 
In ChPT, $O_4$ is represented by 
\begin{equation}
O_4=
\alpha_{\scriptscriptstyle 27} r^{ij}_{kl}({\Sigma \partial^\mu 
\Sigma^\dagger})_i^{\ k}({\Sigma \partial_\mu \Sigma^\dagger})_j^{\ l}
\end{equation}
 at $O(p^2)$ with the tensor $r^{ij}_{kl}$ projecting out the 
 $\Delta S =1$, $\Delta I=3/2$ components, 
 and a large number of $O(p^4)$ operators, 
each with its associated coefficient, 
 constructed from the unitary $\Sigma$-field and the quark-mass matrix $M$
\cite{kametal}. 
A similar representation can be constructed for 
$O'={\bar s}_L\gamma^\mu d_L\ {\bar s}_L\gamma_\mu d_L$, which 
is related to ${\overline K}^0 -K^0$ mixing.  
At $O(p^2)$ this amounts to just changing the tensor $r^{ij}_{kl}$.
$O_4$ and $O'$ are components of the same 27-plet under 
$SU_L (3)$, and therefore the same coefficient
$\alpha_{\scriptscriptstyle 27}$ appears in both $O_4$ and $O'$. 

The matrix element
$\langle \pi^+ \pi^0 |O_4 |K^+ \rangle$ to one loop 
can be obtained by calculating loop diagrams from the $O(p^2)$ operator, 
which gives rise to chiral logarithms
with known coefficients, and which depend on a cutoff $\Lambda$, 
and tree-level contributions from $O(p^4)$ operators. 
Since we do not have enough information to determine
the $O(p^4)$-operator coefficients, we will
estimate the size of one-loop corrections by setting all 
these coefficients to zero and 
choosing $\Lambda$ to be 770 MeV or 1 GeV, which 
is generally believed to be the energy scale 
below which physical effects of more massive 
hadrons can be  absorbed into the $O(p^4)$ coefficients. The
sensitivity of physical quantities to these different 
values of the cutoff is taken as an estimate of the
systematic error associated with the lack of knowledge of 
the $O(p^4)$ coefficients.

\section{Continuum result}
The one-loop calculation of $\langle \pi^+ \pi^0 |O_4 |K^+ \rangle$ in 
ChPT (for $m_\pi =0$) was first undertaken in \cite{a1}.  
Numerically, the real-world result 
for $\langle \pi^+ \pi^0 |O_4 |K^+ \rangle$, 
with $m_u =m_d \neq m_s$, 
and $m_\pi=136$ MeV, $m_K =496$ MeV and $f_\pi =132$ MeV, is \cite{a4}
\begin{eqnarray}
\langle \pi^+ \pi^0 |O_4 |K^+ \rangle\!\!\!\!&=&\!\!\!\!
{{12i \alpha_{\scriptscriptstyle 27}} 
\over 
{\sqrt{2} f_\pi^3}}
\left(m^2_K -m^2_\pi \right)\times \nonumber \\
&&\!\!\!\!\!\!\left( 1+
{
{0.63,\ \Lambda =1\phantom{70}\ {\rm GeV}}
\atop 
{0.36,\ \Lambda =770\ {\rm MeV}}
}\right).\label{rwdecay}
\end{eqnarray}
We see that the one-loop correction is fairly large.

\section{Lattice results}
It is possible to extract a related matrix element 
on a lattice with spatial volume $L^3$ 
from a computation of the correlation function
\begin{eqnarray}
C(t_2 ,t_1)&\!\!\!\!\!\equiv\!\!\!\!\!&
\langle 0|\pi^+ (t_2) \pi^0 (t_2)O_4 (t_1)K^- (0)|0\rangle 
\nonumber \\
&&\!\!\!\!\!\!\!\!\!\!\!\!\!\!\!\!\!\!\!\!\!\!\!\!\!\!\!\!
{\buildrel{\scriptstyle {t_2 \gg t_1 \gg 0}} \over 
\longrightarrow}e^{-E_{2\pi} (t_2-t_1)} e^{-m_K t_1}
\times \label{correl} \\
&&\!\!\!\!\!\!\!\!\!\!\!\!\!\!\!\!\!\!\!\!\!\!\!\!\!\!\!\!
{{\langle 0|\pi^+\pi^0|\pi^+ \pi^0 \rangle \langle \pi^+ \pi^0 
|O_4|K^+ \rangle \langle K^+ |K^-|0 \rangle} 
\over
{\langle \pi^+ \pi^0 |\pi^+ \pi^0 \rangle \langle K^+|K^+ \rangle}
}\,, \nonumber 
\end{eqnarray}
where $\pi^+ (t) =\sum_{\vec x}\pi^+ ({\vec x},t), 
\ \pi^+\equiv\pi^+(0),\ {\rm etc.}$
and $E_{2\pi}$ is the energy of a state with two pions at rest 
in a finite volume. (One also needs kaon and 
two-pion correlation functions.)
Except for $m_K=2m_\pi$, an ``unphysical" matrix element is obtained  
since all external mesons are at rest. Also, 
in current lattice computations \cite{jlqcd}, unphysical masses 
(degenerate quark masses 
which are heavier than real-world ones), 
the quenched approximation and, of course, 
finite volume are used. 
We will use quenched ChPT \cite{bg} to calculate
$\langle\pi^+\pi^0|O_4|K^+\rangle^{quenched}$ to one loop.
The result can then be compared to eq.~(\ref{rwdecay}) to get an
estimate of all these systematic effects.
It should be noted that the coefficients $\alpha_{\scriptscriptstyle 27}$ and 
$\alpha^q_{\scriptscriptstyle 27}$, the tree-level meson decay
constants $f$ and $f_q$ ($q$ for quenched), etc., are in principle
not equal. As for finite volume, we have: 
1. the difference between finite-volume and continuum values for 
operator coefficients are exponentially 
small in $L$ (ESL) \cite{a2}; 2. spatial momentum integrals  
$\int {d^3{\vec k}\over(2\pi)^3}\  f(k^2,m_\pi )$ 
are replaced by discrete sums
${1\over L^3}\sum_{\vec k ={{2\pi\over L}{\vec n}}} f(k^2,m_\pi ), 
{\vec n} \in Z\!\!\!Z^3$ (for periodic boundary conditions). 
If the integrand is regular, the sums are equal to the
continuum integrals with corrections ESL. Otherwise, there are 
additional power corrections in $L^{-1}$ \cite{a3}: 
\begin{eqnarray}
{1\over L^3}
\sum_{{\vec k}\neq 0} {f(k^2,m_\pi )\over k^2} 
&\!\!\!=\!\!\!& 
\int{d^3 k \over (2\pi)^3} 
{f(k^2, m_\pi )\over k^2} \label{fv} \\
&&\!\!\!\!\!\!\!\!\!\!\!\!\!\!\!\!\!\!\!\!%
\!\!\!\!\!\!\!\!\!\!\!\!\!\!\!\!\!\!\!\!\!\!\!
-{0.22578\over L} f(0, m_\pi) -{1\over L^3}{df \over dk^2}(0,m_\pi)
+{\rm ESL}.
\nonumber
\end{eqnarray}
We will not consider ESL corrections. We calculated
the unphysical amplitude for  
degenerate quark masses, in which case $O_4$ and $O'$ do not couple 
to the $\eta'$-meson, and hence there are no contributions
from the ``$\eta'$-parameters" $\delta$ and $\alpha$ \cite{a4}. 

Details of the calculation 
can be found in \cite{a4}.
The diagram in which the two pions produced from the $K^+$-decay
strongly rescatter leads to  power-like finite-volume 
effects. Suppose the strong-interaction vertex acts at $t_s$ with 
$t_1 < t_s < t_2$. The essential part of the expression for this diagram is
$L^{-3} \sum_{\vec k}
\int_{t_1}^{t_2}\!\!\!dt_s\exp{[-2(\sqrt{{\vec k}^2+m^2_\pi}-m_\pi)(t_s-t_1)]}$.
 For $\vec k =0$, this 
gives $(t_2 -t_1)/L^3$ which can be 
resummed into the tree-level result, and thus  produces 
$E_{2\pi}=2m_\pi+1/(2L^3f_\pi^2)$ in the exponent in 
eq.~(\ref{correl}). For $\vec k \neq 0$, we get terms 
$\propto 1/(\sqrt{m_\pi^2+{\vec k}^2}-m_\pi)=
(\sqrt{m_\pi^2+{\vec k}^2}+m_\pi)/{\vec k}^2$.
These give rise to 
sums like eq.~(\ref{fv}), which results in 
$L^{-1}$ and $L^{-3}$ corrections. We will discard any excited-state 
contributions \cite{a4}.

The collected one-loop corrections for the unphysical 
matrix element $\langle \pi^+ \pi^0 |O_4 (0)|K^+ \rangle$
are
\begin{equation}
{m^2_\pi \over (4 \pi
f)^2}\left(-6\log{m^2_\pi \over \Lambda^2}+F(m_\pi L)    
\right)
\label{fdecay}
\end{equation}
relative to the tree-level value ${{24i \alpha_{\scriptscriptstyle
27} m^2_\pi L^3}/{\sqrt{2}f^3}}$ for the full theory, and
\begin{equation}
{m^2_\pi \over (4 \pi f_q)^2}\left( -3\log{m^2_\pi \over
\Lambda_q^2}+F(m_\pi L)\right) \label{qdecay}
\end{equation}
relative to the tree-level value 
${{24i\alpha^q_{\scriptscriptstyle 27} m^2_\pi L^3}/
{\sqrt{2}f^3_q}}$ for the quenched theory.
\begin{equation}
F(m_\pi L)={17.827/{(m_\pi L)}}+{12\pi^2 /{(m_\pi L)}^3}
\end{equation}
is the finite-volume correction. 

\section{Numerical examples}
To one loop in ChPT the physical matrix element 
and the unphysical one from 
quenched lattice computations (after extrapolation to the continuum
limit) are related by
\begin{eqnarray}
{\langle \pi^+ \pi^0|O_4 (0)|K^+ \rangle
}_{phys}=Y\;{\alpha_{\scriptscriptstyle 27}\over
\alpha^q_{\scriptscriptstyle 27}}\left({f_q \over f}\right)^3 \times 
\nonumber \\
{{m_K^2 -m_\pi^2} \over 2M^2_\pi}\;
{\langle \pi^+ \pi^0 |O_4 (0)|K^+ \rangle
}^{quenched}_{unphys} ,\label{qrelat}
\end{eqnarray}
with
\begin{equation}
Y={
{1+{ {\phantom{-} 0.089,\ \Lambda =1\phantom{70}\ {\rm GeV}}\atop {-0.015,\ 
\Lambda=770\ {\rm MeV}} }}
\over
{1+{M^2_\pi \over (4 \pi
F_\pi)^2}\left[-3\log{M^2_\pi \over \Lambda^2_q}+F(M_\pi
L)\right]}}\ . \label{Y}
\end{equation}
The conversion factor $Y$ embodies all the one-loop corrections.  
The $\alpha_{\scriptscriptstyle 27}$ and $f$ ratios 
in the prefactor are not known. We will arbitrarily set them 
equal to one. It remains one of the uncertainties
that cannot be resolved within ChPT.
$M_\pi$ and $F_\pi$ refer to values computed 
on the lattice while $m_K$ and $m_\pi$ refer to real-world 
values. When we apply the formula to 
the lattice data of \cite{a5} 
(in which the mass-squared-ratio prefactor 
was already taken into account), 
we get values which are shown in fig.~1, along with
the original data (for which the error bars are statistical only). 
The error bars on our points come from varying $\Lambda$ and
$\Lambda_q$ independently, and do not contain the statistical errors.
We have  eliminated points with smaller
physical volume or at which 
$M_\pi >770$ MeV $\approx m_\rho$. 
At all points, $Y < 1$ and each 
``corrected" amplitude lies below the corresponding original 
one. The one-loop 
results reduce the discrepancy 
between the lattice data and the experimental result. 
However, one-loop effects are rather substantial, and  two-loop 
corrections can probably not be neglected. 
For more discussion of all uncertainties involved in these
estimates, we refer to \cite{a4}.
%13%%%%%%%%%%%%%%%%%%%%%%%%%%%%%%%%%%%%%%%%%%%%%%%%%%%%%%%%%%%%%%%%%%%%%
% FIGURE 1
%
\vspace*{-2.8cm}
\begin{figure}[htb]
\epsfxsize=1.0 \hsize
\epsfbox{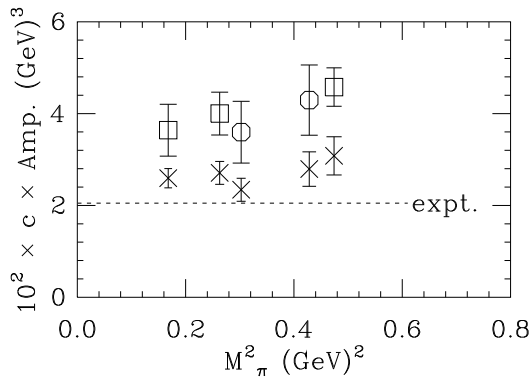}
\caption[]{Open symbols: data from \cite{a5}
(squares: $16^3\times 25$ (or $\times 33$), $\beta=5.7$;
octagons: $24^3\times 40$, $\beta=6$);  crosses:
including the correction factor $Y$.
The constant $c=2\sqrt{2}/(G_F\sin\theta_c\cos\theta_c)$.}
\vspace*{-0.0cm}
%\label{FIG1}
\end{figure}
%%%%%%%%%%%%%%%%%%%%%%%%%%%%%%%%%%%%%%%%%%%%%%%%%%%%%%%%%%%%%%%%%%%%%%%%

A similar relation involving instead the unphysical 
unquenched matrix element is
\begin{eqnarray}
{\langle \pi^+ \pi^0 |O_4 (0)|K^+ \rangle
}_{phys}=X\;
{{m_K^2 -m_\pi^2} \over 2M^2_\pi} \times \nonumber \\
\ \ \ \ \ {\langle \pi^+ \pi^0 |O_4 (0)|K^+ \rangle
}^{full}_{unphys}\ , \label{frelat}
\end{eqnarray}
with
\begin{equation}
X={
{1+{ {\phantom{-} 0.089,\ \Lambda =1\phantom{70}\ {\rm GeV}}\atop {-0.015,\ 
\Lambda=770\ {\rm MeV}} }}
\over
{1+{M^2_\pi \over (4 \pi
F_\pi)^2}\left[-6\log{M^2_\pi \over \Lambda^2}+F(M_\pi
L)\right]}}\ . \nonumber
\end{equation}
$\alpha_{\scriptscriptstyle 27}$ 
and $f$ drop out, unlike in eq.~(\ref{qrelat}). 
At $M_\pi =0.4$ GeV, 
$M_\pi L=8$ and $F_\pi =132$ MeV, $X \approx 0.6$.

\section{Relation to $B_K$}
One-loop results for $B_K$ with $m_u =m_d \neq m_s$ 
can be found in \cite{a4}. The $\eta'$-double 
pole contributes and hence the result depends on $\delta$ and 
$\alpha$ for $m_u =m_d \neq m_s$. 

The ratio
\begin{equation} 
{\cal R}=f_K\; {\langle \pi^+ \pi^0 |O_4 |K^+ \rangle / 
{\ \langle {\overline K}^0 |O'|K^0 \rangle}} 
\end{equation}
in which 
$\alpha_{\scriptscriptstyle 27}$ drops 
out is of some interest. In the physical case, the one-loop 
correction is 60\% to 95\% of the tree-level value 
in magnitude, depending on the values of the cutoff. This calls into question 
the reliability of ChPT in this case. In the unphysical case,
\begin{eqnarray}
{{\cal R}^{full}_{unphys}}&\!\!\!=\!\!\!&{{\cal R}^{quenched}_{unphys}}
=\label{R} \\
&&\!\!\!\!\!\!\!\!\!\!\!\!\!\!\!\!\!\!\!{3i\over\sqrt{2}}
\Biggl[1+
{M^2_\pi \over (4\pi  F_\pi)^2}\left( 3\log{M^2_\pi \over \Lambda^2}+F(M_\pi 
L)\right)
\Biggr]. \nonumber
\end{eqnarray}
The typical magnitude of the one-loop corrections for the data of \cite{a5} 
is 10--15\%.

\smallskip
We would like to thank Claude Bernard and Steve Sharpe for
discussions.  This work was supported in part by the DOE.


\begin{thebibliography}{9}
\bibitem{kametal} J.\ Kambor, J.\ Missimer, D.\ Wyler, 
Nucl. Phys. B346, 17 (1990).
\bibitem{a1} J.\ Bijnens, H.\ Sonoda, M.\ Wise, 
Phys. Rev. Lett. 53 (1984) 2367.
\bibitem{a4} M.\ Golterman, K.C. Leung, Phys. Rev. D56 (1997) 2950.
\bibitem{jlqcd} N.\ Ishizuka for JLQCD, these proceedings.
\bibitem{bg} C.\ Bernard, M.\ Golterman, Phys. Rev. D46 (1992) 853.
\bibitem{a2} J.\ Gasser, H.\ Leutwyler, Phys. Lett. B184 (1987) 83.
\bibitem{a3} M.\ L\"uscher, Comm. Math. Phys. 105 (1986) 153.
\bibitem{a5} C.\ Bernard, A.\ Soni, Nucl. Phys. B (Proc. Suppl.) 17 (1990) 495.
\end{thebibliography}
\end{document}